# Reversible and irreversible magnetocaloric effect in the NdBa$_2$Cu$_3$O$_7$ superconductor in relation with specific heat and magnetization

T. Plackowski [1,2], Y. Wang [1], R. Lortz [1], A. Junod [1], and Th. Wolf [3]

[1] *Département de Physique de la Matière Condensée, Université de Genève,*

*CH-1211 Genève 4, Switzerland*

[2] *Institute of Low Temperature & Structure Research,*

*ul. Okólna 2, 50-422 Wrocław, Poland*

[3] *Forschungszentrum Karlsruhe, Institut für Technische Physik,*

*Postfach 3640, D-76021Karlsruhe, Germany*

**Abstract**

A recently developed technique for measuring the isothermal magnetocaloric coefficient ($M_T$) is applied to the study of a superconducting NdBa$_2$Cu$_3$O$_7$ single crystal. Results are compared with magnetization ($M$) and specific heat ($C$). In the reversible region both $C$ and $M_T$ follow the scaling law of the 3D-xy universality class. The anomalies connected with flux-line lattice melting are visible on $M_T(B)$ curves as peaks and steps, similar to $C(T)$ curves yet with much smaller background. At lower temperature, in the irreversible region the $M_T(B)$ behaviour resembles more that of $M(B)$, exhibiting the "fishtail" effect. Our results confirm that the peculiarities of the phase diagram known from the high temperature superconductor YBa$_2$Cu$_3$O$_7$, e.g. vortex melting, dominance of critical fluctuations and absence of a $B_{c2}$ critical field line, are a common property of RE-123 systems.

## 1. Introduction

In a recent paper [1] we described a method to measure the isothermal magnetocaloric coefficient, defined as:

$$M_T \equiv \left(\frac{\delta Q^\uparrow}{\delta B_a}\right)_T = \left(\frac{dQ^\uparrow/dt}{dB_a/dt}\right)_T \quad (1)$$

where by convention $Q^\uparrow$ is the heat flowing out of the sample and $B_a \equiv \mu_0 H$ the applied magnetic field. The interpretation of this quantity is more straightforward than that of the adiabatic magnetocaloric coefficient $M_S \equiv (dT/dB_a)_S$, since $M_T$ is measured at constant temperature $T$ rather than at constant entropy $S$. $M_T$ is closely related to the magnetization $M$; both quantities share the same units (J/gatT = Am$^2$/gat when expressed per gram-atom, A/m when expressed per volume unit; for Nd-123, one gat occupies 8.25 cm$^3$), and in reversible conditions the following relation holds:

$$M_T = -T\left(\frac{\partial M}{\partial T}\right)_{B_a}. \quad (2)$$

$M_T$ and $C$ can be both considered as generalized specific heats $C \equiv (\delta Q/\delta x_i)_{x_{j\neq i}=const}$: both quantities measure the entropy absorbed (or released) under the effect of an incremental change of an intensive variable $x_i$, here either the magnetic field or the temperature.

In this paper we compare the results of three complementary thermodynamic quantities, $M_T$, $M$, and $C$, for a single-crystal of NdBa$_2$Cu$_3$O$_7$ (Nd-123), which is a member of the "123" family of high-$T_c$ superconductors (HTS). The phase diagram of HTS is particularly rich. In contrast to classic type-II superconductors, there is no true phase transition occurring at the mean-field upper critical field $B_{c2}(T)$ for $B_a > 0$. Below $B_{c2}(T)$ a fluid phase of vortices occupies a large part of the phase diagram. The only true phase transition occurs on the melting line, $B_m(T)$, where the vortex liquid freezes into a solid upon lowering the temperature or the field. This transition manifests itself by a sudden drop of resistivity to zero and the onset of a finite shear modulus of vortex matter, which corresponds to a first- or second-order thermodynamic phase transition, depending on whether the vortex solid is periodic or glassy [2-16].

The melting transition in specific heat measurements has only been observed in Y-, Eu-, and Dy-123 compounds up to now [9-19]. In recent years, significant improvements were achieved in the crystal growth technology of Nd-123 [20, 21], allowing this superconductor to be in turn studied with respect to its vortex properties and critical fluctuations using calorimetric techniques. Therefore, the purpose of the current paper is both to compare the superconducting properties of Nd-123 with the other members of the 123 family, and to test the usefulness of the recently developed isothermal magnetocaloric technique to study the thermodynamics of HTS.

## 2. Experimental

Nd-123 single crystals were grown from flux in a $SnO_2$ crucible using $Nd_2O_3$, $BaCO_3$ and CuO with purities of 4N to 5N. The atmosphere in the furnace was kept at 60 mbar air to avoid Nd substitution on the Ba site. Oxidation was carried out in 1 bar $O_2$ in the temperature range from 500 to 310°C during 400 hours and in addition in 963 bar $O_2$ at 320°C during 232 hours.

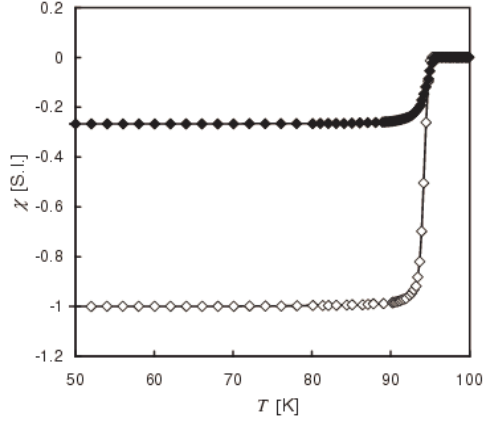

**Figure 1.** ZFC (lower curve) and FC (upper curve) static susceptibility of the Nd-123 single crystal measured at $B_a = 2$ mT.

Figure 1 shows the superconducting transition of the sample measured in a SQUID magnetometer using zero field cooled (ZFC) and field cooled (FC, Meissner-Ochsenfeld) magnetization. The onset of the FC transition occurs at 95.5 K, the transition width (10% - 90%) is 1.5 K. The magnetization was measured in a Quantum Design MPMS-5 magnetometer. The specific heat was measured from 30 to 120 K in a continuous-heating adiabatic calorimeter operating at 0 – 14 T [22]. The heating rate was ~15 mK/s.

The isothermal magnetocaloric coefficient was measured using a heat-flow calorimeter [1]. In this method, the sample is linked to a heat sink by means of a sensitive heat-flow meter of high thermal conductance $\kappa$. The heat flux was sensed by a miniature single-stage Peltier cell with a sensitivity of $A = 0.45$ V/W at room temperature (RT), and $A = 0.08$ V/W at liquid nitrogen temperature ($LN_2$) [23]. The sensitivity increases slightly with magnetic field, e.g. by 11 % at RT for a field of 13 T normal to the cell surface. The thermal conductance of the cell varies little, from $\kappa = 28$ mW/K at RT to 35 mW/K at $LN_2$. The sample was thermally anchored to the top plate of the cell, made of 0.5 mm thick alumina. The bottom of the heat-flow meter was soldered to the heat sink, a massive temperature-controlled copper block. An in-field calibrated Pt thermometer was attached to the sink. The calorimeter was evacuated to $10^{-6}$ hPa and placed in the gas flow of a variable-temperature insert, inside the bore of a 14/16 T superconducting magnet.

For measurements of the isothermal magnetocaloric coefficient the sink temperature was kept constant, while the magnetic field was swept at a constant rate $dB_a/dt$. In this way the magnetocaloric effect was studied in quasi-isothermal conditions. The $M_T$ measurements can be performed upon increasing or decreasing the field. The empty cell is non-magnetic and does not contribute to the signal.

## 3. Results

### *3.1 Magnetization*

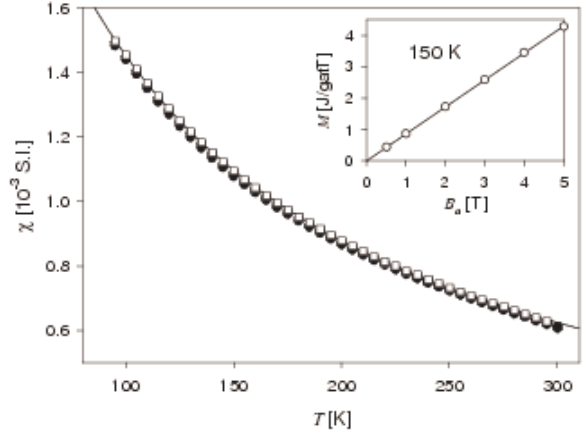

**Figure 2.** Normal-state static susceptibility versus temperature for $B_a$ = 0.5, 1, 2, 3, 4 and 5 T. The insert shows the linear field dependence of the magnetization at 150 K.

The normal-state susceptibility is shown in figure 2. All curves follow a Curie-Weiss law $\chi = \chi_0 + C_{CW}/(T - \Theta_{CW})$ with $\chi_0 = -3 \times 10^{-5}$, $C_{CW}$ = 0.235 K and $\Theta_{CW}$ = −58 K. The normal-state magnetization is linear in the magnetic field (inset of figure 2). In these conditions, equation (2) predicts for the isothermal magnetocaloric coefficient $M_T = T(B_a/\mu_0)[C_{CW}/(T-\Theta_{CW})^2]$.

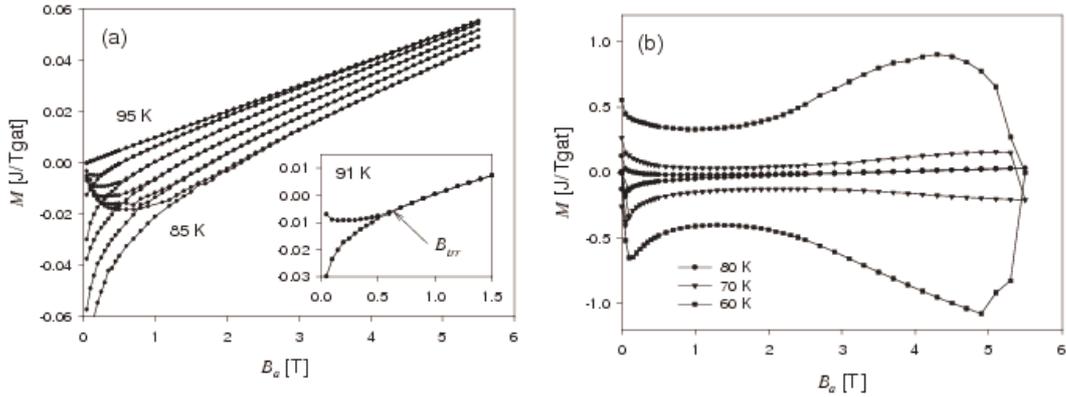

**Figure 3.** Magnetization versus the magnetic field along the *c*-axis in the superconducting state: (a) reversible region, from 85 to 95 K by 2 K increments. The inset shows the working definition of the irreversibility field, $B_{irr}$; (b) irreversible region.

Figure 3 shows the magnetization versus magnetic field in the superconducting state for $B_a$ applied along the *c*-axis. The sample shows high reversibility at high temperature (figure 3a). The irreversibility field, $B_{irr}$, is determined from the point where the $M(B_a)$ loop closes (inset of figure 3a). Below $T$ = 88 K, the hysteresis vanishes very slowly so that this definition of $B_{irr}$ is no longer robust. Figure 3b presents magnetization loops in the irreversible region exhibiting a fishtail effect with a minimum hysteresis around $B_a = 1.5$ T.

*3.2 Specific heat*

Figure 4a shows the specific heat as a function of temperature in different magnetic fields from 0 to 14 T. The field is again applied along the *c*-axis. The inflexion point of *C/T* at $B_a = 0$ gives $T_c = 95.5$ K. A slight increase of the specific heat with the magnetic field due to the magnetic contribution of $Nd^{3+}$ ions can be seen at $T > T_c$. Figure 4b shows the variation of the specific heat with the field $\Delta C(B_a, T)/T \equiv [C(B_a) - C(0)]/T$. The peak features observed at $T < T_c$ between 3 and 12 T are characteristic of first-order vortex-melting transitions, as observed in Y-123, Dy-123 and Eu-123 [9-19].

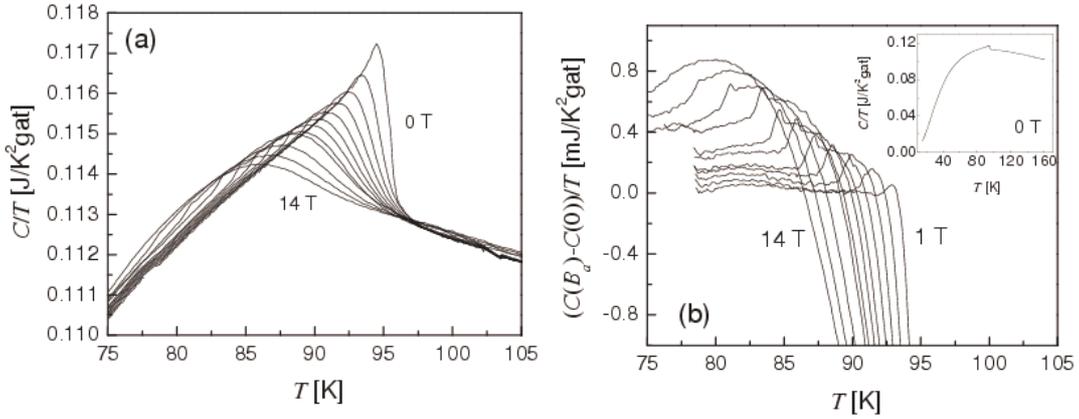

**Figure 4.** (a): Specific heat *C/T* versus temperature in the vicinity of the superconducting transition for various magnetic fields along the *c*-axis, from right to left $B_a = 0, 1, 2, 3, 4, 5, 6, 7, 8, 10, 12$ and 14 T. (b): Variation of the specific heat with the field $[C(B_a) - C(0)]/T$ versus temperature at $T < T_c$, showing details of the vortex-lattice melting transition, from bottom to top $B_a = 1, 2, 3, 4, 5, 6, 7, 8, 10, 12$ and 14 T. Insert: *C/T* in zero field over a wide temperature range.

For HTS, thermal fluctuations of the order parameter in the vicinity of $T_c$ cause deviations from the mean-field theory. The latter would only predict a specific heat jump at $T_c$. The description of fluctuations in the critical region can be obtained from the asymptotical form of the singular part of the free energy [24]:

$$F_{fl} = -k_B T \xi^{-D} f_1(\xi/a) \qquad (3)$$

with $k_B$ Boltzmann's constant, $\xi$ the correlation length, $D$ the dimension of fluctuations, $a = (\Phi_0/B_a)^{1/2}$ the magnetic length, $\Phi_0$ the quantum of flux, and $f_1$ some function of a single variable. Assuming further a divergence of the form $\xi(t) = \xi^{\pm} |t|^{-\nu}$, where $t \equiv (T - T_c)/T_c$ and $\nu$ is a critical exponent, one finds that in the limit $\xi/a \to 0$ the singular part of the specific heat diverges at $T_c$ with an exponent $\alpha = 2 - D\nu$ (see e.g. Ref.[25-27]):

$$\frac{C_{fl}^{\pm}}{T} = -\frac{\partial^2 F_{fl}}{\partial T^2} \approx T \frac{A^{\pm}}{\alpha} |t|^{-\alpha} \quad (4),$$

where "+" and "−" refer to $T > T_c$ and $T < T_c$, respectively. In practice the following equivalent approximation is preferred when $|\alpha| \ll 1$. It leads to a more realistic non-singular background in the fitting procedure:

$$\frac{C_{fl}^{-}}{T} \approx A^{+}(J - \ln|t|), \quad \frac{C_{fl}^{+}}{T} \approx -A^{+} \ln t \quad (5).$$

There is evidence that the transition of HTS belongs to the same universality class as the superfluid transition of $^4$He, i.e. the 3D-xy model characterized by $\nu \cong 0.67$, $\alpha \cong -0.013$, and a specific-heat jump $J \equiv (A^{-}/A^{+} - 1)/\alpha \cong 4.0$ [28].

The specific heat of our sample in zero field may be fitted with the function $C(T) = C_{bg}(T) + C_{fl}^{\pm}(T)$, where the regular background $C_{bg}(T)$, mostly due to the lattice contribution, is represented by two Einstein modes in the vicinity of $T_c$. Figure 5 shows data and fit from 75 to 115 K. The model describes well the transition in zero field, except for the jump $J$ which is larger, possibly indicating amplitude fluctuations. The main parameters are listed in Table I, together with data for other HTS given for comparison purposes.

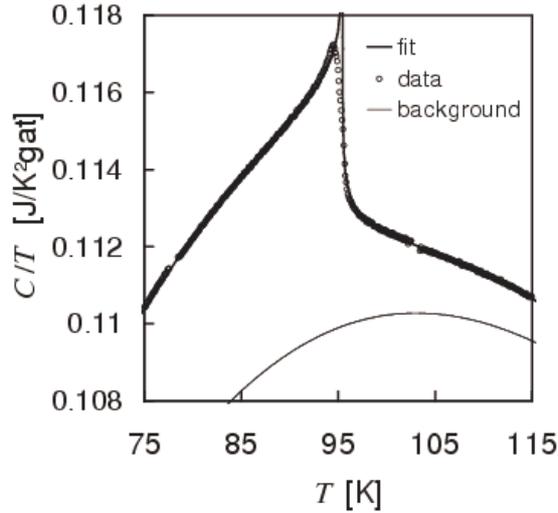

**Figure 5.** Specific heat $C/T$ versus temperature at zero field. Symbols: measured data, thick solid line: 3D-xy fit, thin solid line: regular background.

|  | NdBa$_2$Cu$_3$O$_7$ | YBa$_2$Cu$_3$O$_7$ [29] | DyBa$_2$Cu$_3$O$_7$ [29] | EuBa$_2$Cu$_3$O$_7$ [29] |
|---|---|---|---|---|
| $T_c$ (K) | 95.5 | 88.3-87.8 | 90.3 | 94.4 |
| $A^+$ (mJ/K$^2$gat) | 0.69 | 0.70-0.72 | 0.68 | 0.78 |
| $J$ | 5.8 | 5.9-6.1 | 6.3 | 5.6 |
| $\Delta C/T_c$ (mJ/K$^2$gat) | 4.0 | 4.1-4.4 | 4.3 | 4.4 |

Table I. Parameters of the fits according to equation (5) for Nd-123 and other HTS.

### 3.3 Isothermal magnetocaloric effect

The isothermal magnetocaloric coefficient is shown in Figure 6a-d. Figure 6a presents curves obtained just above $T_c$ (black curves) or just below (colour curves). Data for increasing (thicker lines) and decreasing field (thinner lines) exhibit good reversibility in this temperature range. The normal-state curve is a straight line with a slope of 5.8×10$^{-3}$ J/gatT$^2$.

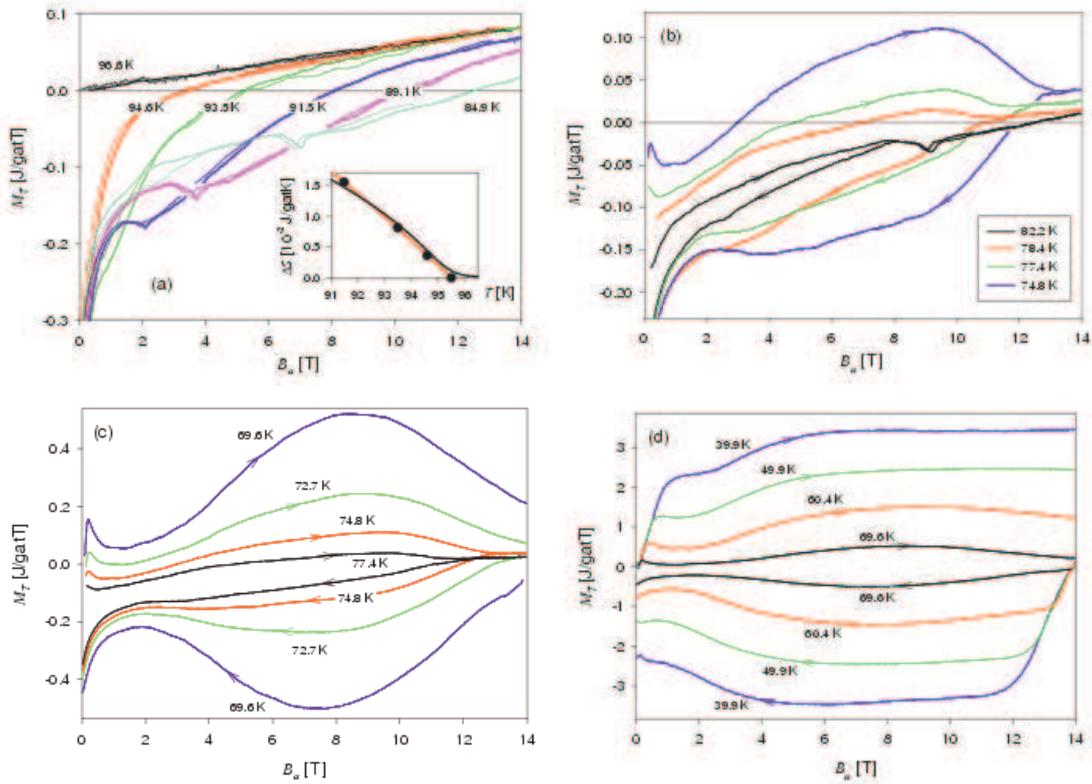

**Figure 6.** Isothermal magnetocaloric coefficient versus magnetic field. Thicker and thinner lines in panel (a) denote data collected upon increasing and decreasing field, respectively. The direction of the variation of the field is indicated by arrows in the other panels. The inset in panel (a) shows the difference between the entropy in the normal state and that in the superconducting state obtained by integrating the $M_T(B_a)/T$ curves (black points) or the $C(T,B_a)/T$ curves (black line). The red curve is a fit of the two-fluid model (see text for details).

The reversible $M_T(B_a)$ curves taken just below $T_c$ (red and green lines) may be used to estimate the superconducting condensation energy $E_c \equiv (V/2\mu_0)B_c^2(0)$, where $B_c(T)$ is the thermodynamic critical field at a temperature $T$. By integrating the curves up to $B_{c2}$, one can measure $\Delta S(T)$, the entropy difference between the normal and superconducting state:

$$\Delta S(T) = -\frac{1}{T}\int_0^{B_{c2}(T)}[M_T(T,B_a) - M_T^{para}(T,B_a)]dB_a \qquad (6)$$

Here $M_T^{para}$ denotes the normal-state paramagnetic component of $M_T$, which can be measured just above $T_c$ and easily extrapolated. As we only consider here a narrow temperature range, its temperature dependence may even be neglected. At $T = 91.5$ K, $B_{c2}(91.5\text{ K}) > 14$ T so that the $M_T(B_a)$ data require a reasonably short extrapolation. The temperature dependence of $\Delta S(T)$ thus obtained from the magnetocaloric effect is presented in the inset in figure 6a. It varies almost linearly with $T$ at a rate of $-3.5$ mJ/gatK$^2$. This linearity is a mean-field property which has to break down close to $T_c$. It can be accounted for by using the two-fluid model $B_c(T) = B_c(0)(1-(T/T_c)^2)$. In this approximation, one has

$$\Delta S(T) = \frac{4E_c(0)}{T_c}\frac{T}{T_c}\left[1-\left(\frac{T}{T_c}\right)^2\right]. \qquad (7)$$

The slope at $T_c$ is $d\Delta S/dT = -8E_c(0)/T_c^2$. Together with our experimental determination, we obtain $E_c(0) = 4.0$ J/gat, a value quite close to $E_c(0) = 3.8$ J/gat estimated from reversible magnetization for Y-123 [30]. Our value for the condensation energy corresponds to $B_c(0) = 1.1$ T, or within the same model $\Delta C/T_c = 8E_c/T_c^2 = 3.5$ mJ K$^{-2}$ gat$^{-1}$ for the specific-heat jump

On the dark blue line measured at 91.5 K, still in the reversible region, an anomaly shows up at $B_a \sim 2$ T. This anomaly is associated with the vortex-melting transition. Its location in $B_a$-$T$ coordinates agrees well with that found in the $C$ measurements. As seen in the next curve at lower temperature (magenta, $T = 89.1$ K) the anomaly shifts toward higher fields and becomes sharper. On the last curve (light blue, $T = 84.9$ K) irreversibility appears below 4 T where the magnetization and demagnetization curves move apart.

This irreversibility becomes increasingly evident at lower temperature (figure 6b). For $T = 82.2$ K (black curve) the irreversibility vanishes at 8 T, just below the melting transition. At still lower temperature, the irreversibility field catches up with the melting transition, then hides or opposes it. Next curves (red, green and dark blue) no longer show any melting anomaly. The hysteresis becomes gradually more important, while the irreversibility field ($B_{irr}$) keeps increasing.

The last curve in figure 6b (T = 74.8 K) shows a new feature at low fields: a minimum in the irreversibility near B ~ 2 T. Figure 6c presents the evolution of this fishtail-like shape, which becomes more pronounced at lower temperature without much shifting on the field scale. This feature strongly resembles the fishtail effect observed in magnetization loops. Indeed, they have a common origin, as detailed in the next section.

Figure 6d shows the isothermal magnetocaloric effect at the lowest investigated temperatures. The absolute values of $M_T$ are clearly increasing, but the fishtail effect gradually fades out, whereas $M_T(B)$ becomes less field dependent. Note, that at the beginning of each field sweep (both at $B = 0$ and after the field reversal at the maximum field) a new behaviour sets in. All curves start from $M_T = 0$, after which $|M_T|$ increases linearly along a common line, before reaching a temperature-dependent saturation value. Again, we refer to next section for further details.

## 4. Discussion

### 4.1 Scaling: the 3D-xy model

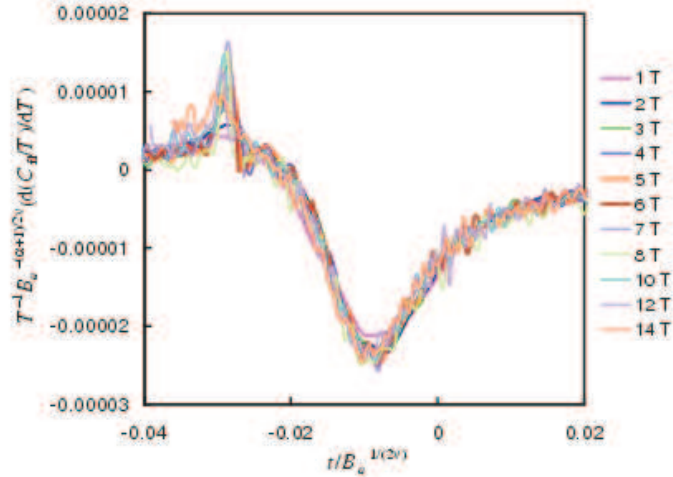

**Figure 7.** 3D-xy scaling plot of the derivative of the specific heat $C_{fl}/T \equiv (C - C_{bg})/T$ versus $T$ and $B_a$ according to equation (9). The fields from 1 to 14 T are those shown in Fig. 4. Units are Tesla, Joule/gat and Kelvin.

In $B_a > 0$, equation (3) leads to the following scaling property for the singular part of the specific heat near $T_c$ [26, 27]:

$$\frac{C_{fl}}{T} = TB_a^{\frac{D}{2}-\frac{1}{\nu}} f_2\left(\frac{t}{B_a^{1/2\nu}}\right), \qquad (8)$$

where $t/B_a^{1/2\nu}$ is one of the forms of the variable $\xi/a$ and $f_2$ is some scaling function. Working with the derivative of $C/T$ is preferred as it dramatically changes the ratio between the singular part of the specific heat and the background, which has zero slope in the vicinity of $T_c$ (see inset of figure 4), in favour of the former:

$$\frac{\partial C/T}{\partial T} \approx \frac{\partial C_{fl}/T}{\partial T} = TB_a^{\frac{D}{2}-\frac{3}{2\nu}} f_3\left(\frac{t}{B_a^{1/2\nu}}\right), \quad (9)$$

where $f_3$ is another scaling function.

Figure 7 shows the corresponding scaling plot in magnetic fields up to 14 T. The same background curve $C_{bg}(T)$ and the same critical temperature are chosen as for the zero-field fit (figure 5). The curves for different fields $B > 1$ T collapse for the critical exponent $\nu = 0.69 \pm 0.03$. The sharp positive peaks at $T < T_c$ correspond to the vortex-melting transition, which occurs for a constant value of the scaling variable, as found for other HTS [22 and references therein].

A similar scaling approach can be applied to magnetocaloric data. The scaling relation takes the same form as for the temperature derivative of the magnetization:

$$M_{T,fl} = -T\frac{\partial S_{fl}}{\partial B_a} = \frac{\partial^2 F_{fl}}{\partial B_a \partial T} = -T\frac{\partial M_{fl}}{\partial T} = -T^2 B_a^{\frac{D}{2}-\frac{1}{2\nu}-1} f_4\left(\frac{B_a}{t^{2\nu}}\right), \quad (10)$$

where $B_a/t^{2\nu}$ is one of the forms of the variable $\xi/a$ and $f_4$ is an unknown scaling function; note that in the above expressions the $T$ prefactor can be considered as equal to $T_c$. A non-singular background arising from the normal-state magnetization $M^{para} = \chi(T)B_a/\mu_0$ must be subtracted first (see figure 2). This correction relative to the singular part of the magnetocaloric coefficient turns out to be much smaller than the phonon background relative to the singular part of the specific heat, so that it is not necessary to go to the next higher-order derivative. In the reversible region, the paramagnetic contribution to the magnetocaloric effect $M_T^{para}(B_a,T)$ can be calculated and reliably extrapolated based on the normal-state susceptibility, using equation (2):

$$M_T^{para} = \frac{B_a T}{\mu_0}\frac{d\chi^{para}}{dT} \quad (11)$$

Equation (11) is verified at $T = 96.6$ K where both $M_T^{para}$ (figure 6a) and $\chi^{para}$ (figure 2) are measured.

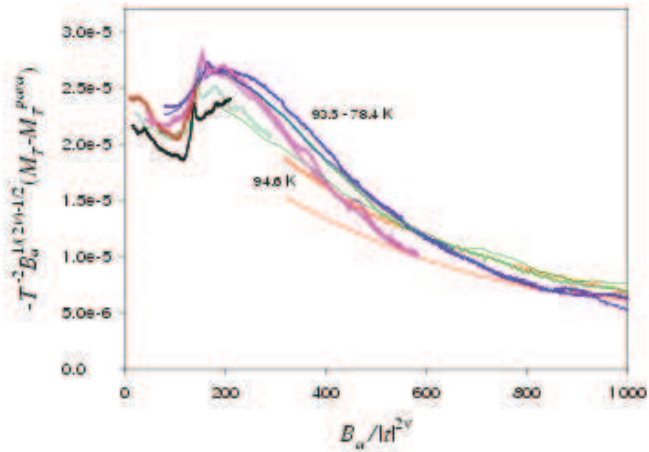

**Figure 8.** 3D-xy scaling plot of the superconducting contribution to the isothermal magnetocaloric coefficient, $M_{T,fl} = M_T - M_T^{para}$, versus $B_a$ and $T$ according to equation (10). Units are Tesla, Joule/gat and Kelvin.

The singular part of the magnetocaloric coefficient $M_{T,fl} = M_T - M_T^{para}$ is presented in the scaling plot of figure 8. Same values of $\nu$ and $T_c$ are used as for the scaling of the specific heat. The data for $B < 0.5$ T are omitted for $T = 94.6$ K. For $78.4 \leq T \leq 84.9$ K, where the irreversibility is relatively weak, we plot the average of the results for increasing and decreasing field. The scaling law is reasonably well obeyed, including the position of the vortex-lattice melting anomaly, except for the temperature region close to $T_c$. This is usual: in zero field, the full divergence of the specific heat, magnetocaloric effect, etc. is never observed since the divergence of the correlation length is cut off by the sample size for an ideal crystal, and by inhomogeneities in real crystals. Scaling holds when finite-size effects are no longer dominant.

*4.2. Vortex-lattice melting*

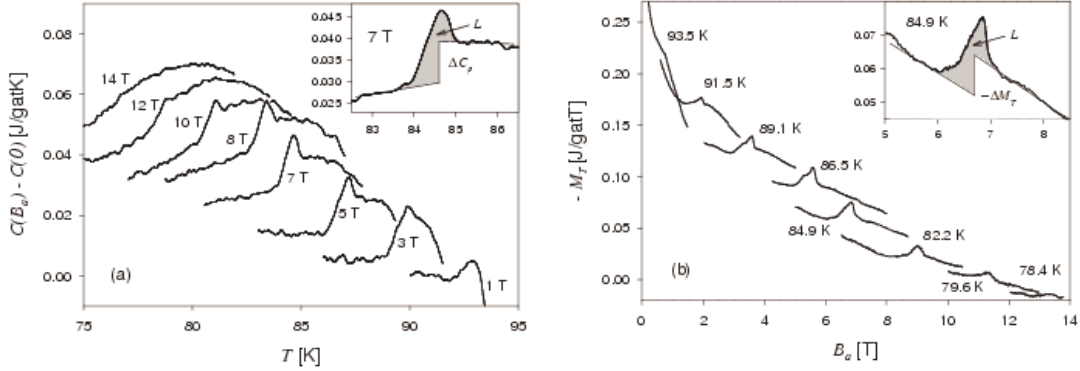

**Figure 9.** Anomalies at the vortex melting transition detected by (a) the specific heat and (b) the isothermal magnetocaloric effect. The insets show how the amplitude of the step ($\Delta C = 9.5 \pm 1$ mJ/gat K, $\Delta M_T = -12 \pm 1$ mJ/gat T) and the latent heat ($L = 8 \pm 1$ mJ/gat in both cases, grey area) are estimated at $T \sim 84.9$ K and $B \sim 7$ T.

The anomalies associated with vortex-lattice melting appear on both $C(T)$ and $M_T(B_a)$ curves (figure 9). In both cases they consist of a superposition of a step and a peak. In some cases the precise determination of the step height and peak area is difficult because of experimental broadening and noise. Nevertheless, it is possible to state that the step height remains approximately constant up to the highest fields or lowest temperature (in average $\Delta C = 10 \pm 2$ mJ/gatK and $\Delta M_T = -12 \pm 2$ mJ/gatT), where it fades out. On the other hand, the best defined peaks are observed in the middle of the range. An example is given in the insets of in figure 9, where numerical values of $\Delta C$, $\Delta M_T$ and $L$ could be determined with better precision.

Since the specific heat and the magnetocaloric effect are thermodynamic quantities, their results are not independent. For a reversible second-order transition, the discontinuity in the slope of the equilibrium magnetization and the specific heat jump obey Ehrenfest's relation:

$$\frac{\Delta(C/T)}{\Delta(\partial M/\partial T)} = -\frac{dB_{tr}}{dT} \quad (12).$$

Together with equation (2) one obtains a similar relation involving the isothermal magnetocaloric effect:

$$\frac{\Delta C}{\Delta M_T} = \frac{dB_{tr}}{dT} \quad (13).$$

At 7 T, the measured specific-heat step $\Delta C$ is $9.5 \pm 1$ mJ/Kgat. Using $dB_m/dT = -0.82$ T/K from the fit of melting line, we find $\Delta M_T = -11.6$ mJ/Tgat. This is in excellent agreement with the measured value $\Delta M_T = -12 \pm 1$ mJ/Tgat.

Turning to the first-order melting transition, the latent heat $L$ can be obtained both from $C$ and $M_T$ by integrating the area under the peaks, as shown in figures 8a and 8b:

$$L = \int C(T)dT = -\int M_T(B)dB \quad (14)$$

We obtain the same value $L = 3.4 \pm 0.3$ mJ/gat from either integral at $T \cong 84.9$ K and $B \cong 7$ T. Using the Clausius-Clapeyron relation we can also estimate the expected value of the reversible magnetization jump on the melting line, $\Delta M = -L/(T_m dB_m/dT) = 4.9 \times 10^{-5}$ J/gatT. Unfortunately, the anomaly in the magnetization could hardly be detected, a problem commonly encountered and associated with surface barriers [31].

### 4.3. Irreversible region

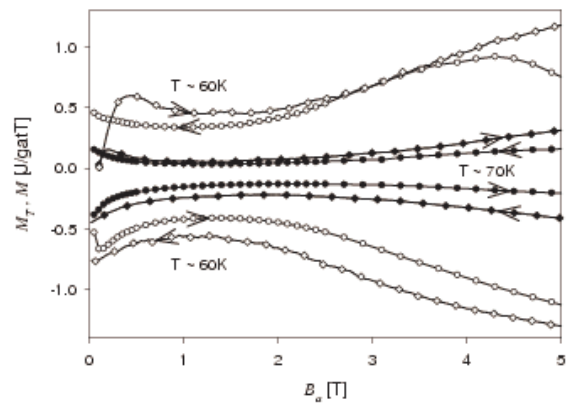

**Figure 10.** Comparison of the irreversible behaviour of the isothermal magnetocaloric coefficient $M_T$ (diamonds) and the magnetization $M$ (circles). Data collected at $T = 69.6$ K for $M_T$ and $T = 70$ K for $M$ are denoted by closed symbols, data collected at $T = 60.4$ K for $M_T$ and $T = 60$ K for $M$ are denoted by open symbols.

Below T = 75 K both the isothermal magnetocaloric coefficient and the magnetization are dominated by irreversibility caused by vortex pinning. They are no longer related by equation (2). However, their behaviour still looks similar. $M_T$ and $M$ are presented on a common plot (figure 10) showing examples measured close to 70 K and 60 K. As seen, both quantities are not only qualitatively but also quantitatively related.

The similarity in the irreversibility of $M_T$ and $M$ may be explained as follows. In the case of magnetization, the irreversibility comes from different distributions of vortex density across the sample upon increasing or decreasing the field (as described by e.g. the Bean or Kim-Anderson models [32, 33]). On the other hand, the origin of the irreversibility in the magnetocaloric effect lies in the friction of vortices against pinning centers. Friction can only result in a heat release $\delta Q^\uparrow > 0$, irrespective of the direction of the field sweep. Thus, the sign of the $M_T$ coefficient, which is defined as $\delta Q^\uparrow / \delta B_a$, is determined by the sign of the field change $\delta B_a$ during the sweep. On the contrary, during a reversible process, heat is released or absorbed depending on the sweep direction, cancelling the sign of $\delta B_a$, so that the sign of the $M_T$ coefficient finally does not depend of the sweep direction.

The similarity between the absolute values of $M_T$ and $M$ originates from the fact that they are both governed by the same material parameter, the critical current density $J_c$. On one hand, $J_c$ determines the gradient of the induction field across the sample $\nabla B_{in} = \mu_0 J_c$, the average induction field in the sample $\langle B \rangle$, and finally the magnetization $M = (\langle B \rangle - B_a)/\mu_0$. On the other hand, $J_c$ is related to the pinning force $F_p = \Phi_0 J_c$, which is obviously the parameter which determines the amount of heat released by vortices as they penetrate the sample. Therefore, the hysteresis defined as $\Delta M = M^- - M^+$ and $\Delta M_T = M_T^+ - M_T^-$ (superscripts denote the direction of the sweep) are both measures of the same $J_c$.

At the beginning of each field sweep the irreversible magnetocaloric curve starts from $M_T = 0$, even after a sweep reversal. This behaviour, which contrasts with that of magnetization, is due to the fact that the sweep reversal causes all vortices to stop; still vortices do not produce heat, whereas their magnetization remains. When the field reverses, the number of moving vortices gradually increases until the induction profile within the sample is fully reversed; $M_T$ reaches then its saturation value determined by $J_c$.

A more detailed description of these processes will be given elsewhere [34]. We point out a property of the magnetocaloric irreversibility visible at low temperature, namely its linear dependence on temperature (figure 11). Here, $\Delta M_T^{max}(T)$ is defined as the maximum value of $\Delta M_T(T, B_a)$ for a given temperature. According to the considerations given above, we expect that $\Delta M_T^{max}$ is proportional to the maximum value of $J_c$. Below ~72 K it varies with temperature with a slope of -0.19 J/gatT$^2$.

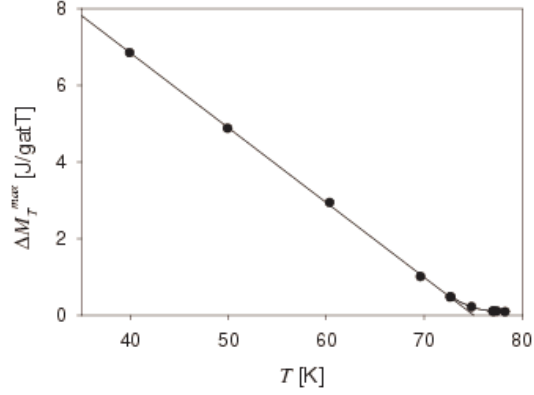

**Figure 11.** Maximum hysteresis of the magnetocaloric effect $\Delta M_T^{\max}(T)$ versus temperature, reflecting the maximum critical current density $J_c(T)$.

*4.4 Phase diagram*

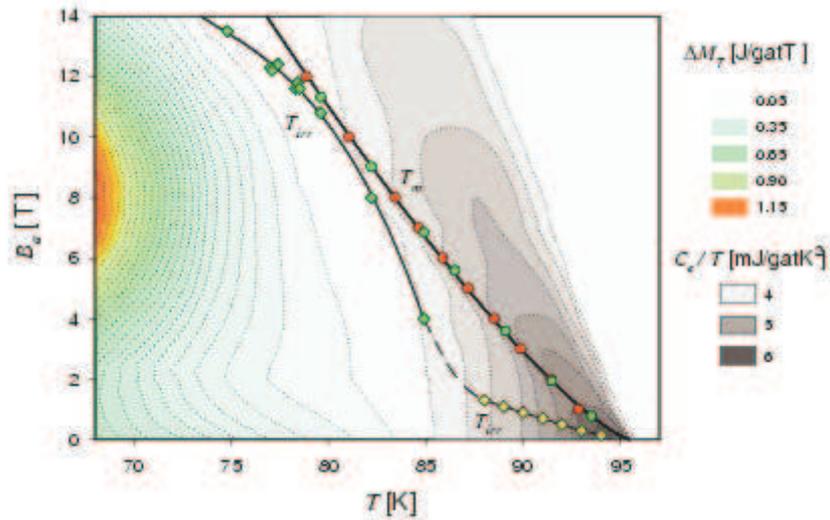

**Figure 12.** Phase diagram of the NdBa$_2$Cu$_3$O$_7$ single crystal in the $B_a$-$T$ plane. Circles: transitions on the flux-line lattice melting line $T_m$ observed either by specific-heat (red) or by magnetocaloric runs (green). Diamonds: points where the hysteresis vanishes, either in magnetocaloric (green) or magnetization loops (yellow), defining the irreversibility line $T_{irr}$. The dash-dotted line connecting both parts of the $T_{irr}$ line is speculative. The colour-scale contour plot on the left represents the magnetocaloric hysteresis $\Delta M_T = M_T^+ - M_T^-$. The gray-scale contour plot on the right represents the excess specific heat $C_e/T$ with respect to the lattice background in the vicinity of the superconducting transition.

The present data collected for the NdBa$_2$Cu$_3$O$_7$ single crystal using specific heat (*C*, red symbols), isothermal magnetocaloric effect ($M_T$, green symbols), and magnetization experiments (*M*, yellow symbols) are summarized in the $B_a$-$T$ phase diagram shown in figure 12. The position of peaks and steps in the *C* and $M_T$ curves defines the melting line $T_m$. The irreversibility line $T_{irr}$ is obtained from *M* at low fields, and from $M_T$ at high fields. The criterion adopted for the onset of irreversibility is the resolution of the respective measurements. The melting line ends where it meets the irreversibility line at high field, as could also be seen in figure 6b. The detrimental effect of hysteresis on melting is also apparent in the fact that the larger the distance between $T_{irr}$ and $T_m$, the sharper the first-order peak in $C$ or $M_T$ on the melting line. This is a consequence of the disorder introduced in the vortex lattice by pinning. The melting line may be fitted by the function $B_m = B_0(1 - T_m/T_c)^{2n}$, $B_0 = 114$ T and $T_c = 95.5$ K and $n = 0.64$. The latter value is close to the 3D-xy exponent $\nu = 0.67$. Thus, this fit shows that melting is driven by the amplitude of critical fluctuations. These parameters are similar to those obtained for RE-123, RE = Y, Dy, and Eu [15].

The line of first-order melting transitions terminates on a lower tricritical point $B_{cr}^{lo} \approx 3$ T and an upper tricritical point $B_{cr}^{up} \approx 10$ T. It is continued by second-order transitions at both ends, which themselves fade away at still lower or higher fields. At low field, the periodicity of the vortex lattice is broken by pinning on widely spaced defects such as twin boundaries; accordingly, twin-free crystals have shown lower values of $B_{cr}^{lo}$. At the other end, high fields tend to decouple the CuO superconducting planes, making them more susceptible to disorder, so that the periodicity is again easily broken. Oxygen vacancies give an example of a source of such short-range disorder; accordingly, fully oxygenated Y-123 and Dy-123 crystals have shown immeasurably high values of $B_{cr}^{up}$ [15, 35]. Only at intermediate fields the interaction with pinning centers can be neglected. The remaining repulsive magnetic interaction between vortices ideally gives rise then to the ordered Abrikosov lattice, the symmetry of which changes upon melting, causing the transition to be of first order.

The colour-scale contour plot representing the magnetocaloric hysteresis $\Delta M_T = M_T^+ - M_T^-$ maps the distribution of critical currents. The island centered on 8 T at low temperature (red area in figure 12) corresponds to the maximum of the fishtail effect shown in figure 6c. The irreversibility line could not be determined by a single type of experiment, owing to limitations of the magnetometer and to the large and steep background that develops as $T$ approaches $T_c$ in $M_T$ runs (figure 6a). Determinations at high and low field are connected by a speculative dash-dotted line in figure 12. Note that this construction continues the shape of the iso-$\Delta M_T$ lines determined at lower temperature, which show a valley extending horizontally along the $B_a \approx 2$ T line.

The magnetization of Nd-123 crystals from the same source was investigated at high fields by means of a vibrating sample magnetometer (VSM) in Ref. [36]. Values of the irreversibility field at 77 K found between 10.1 and 13.4 T for crystals with high oxygen content are consistent with the value 12.5 T determined by the magnetocaloric effect in this work. A close inspection of the VSM irreversibility lines shows that they also reveal a valley-like feature, similar to that found in the present work, albeit much weaker. This difference may result from ageing effects. As shown by figure 1 of Ref. [36], the fishtail effect becomes more pronounced with time when the crystal is stored at room temperature.

In the region above the $T_m$ line, the magnetocaloric effect is dominated by reversible contributions which reflect superconducting correlations slowly dying out (figure 6a). No sharp upper critical field line $B_{c2}$ can be evidenced experimentally. In accordance with specific heat data, which show that the sharp λ-anomaly at $T_c$ in zero field progressively flattens out in a field, the transition at $B_{c2}$ is replaced by a smooth crossover. This can be understood in the framework of finite-size effects where the fluctuations length scale is limited by the magnetic length or the inter-vortex distance [37, 38]. To visualize this behaviour, contours showing the value of the electronic specific heat ($C_e/T$) are shown in gray in figure 12 (not to be confused with colour contours on the left, which represent hysteresis). These contours qualitatively reflect lines where $\xi/a$ is constant in the 3D-xy model.

This phase diagram is qualitatively the same as that of Y-123 [12, 39], although on a quantitative level the hysteresis can vary considerably between samples. This similarity shows that the rich physics resulting from the interplay of vortices, pinning and critical fluctuations is not a peculiarity of one RE123 system, but rather is a general property.

## 5. Summary

In this paper we have studied thermodynamic properties of a single crystal of the $NdBa_2Cu_3O_7$ superconductor. We have presented the measurements of three closely related quantities: magnetization, specific heat and isothermal magnetocaloric coefficient. The technique to measure the latter quantity was recently developed [1] and was applied here to a HTS for the first time. It was shown that it is a useful tool to study this class of materials. $M_T$ owns some features of the specific heat – it measures the variation of the entropy versus field, whereas $C$ measures it versus temperature. Both are second derivatives of the free energy state function. On the other hand $M_T$ owns some features of the magnetization – both $M_T$ and $M$ share same physical units. Moreover, in the case of superconductors they are both composed of two contributions, a reversible one (due to field dependence of the free energy) and an irreversible one (due to the dissipation that accompanies vortex motion). A significant advantage of $M_T$ as compared to $C$ is the absence of phonon contribution, which usually dominates the specific heat of HTS close to $T_c$. Paramagnetism may add a background contribution to $M_T$ (as well as to specific heat), but this background is small and can easily be separated.

We showed that it is possible to use $M_T$ to evaluate the condensation energy. It was also demonstrated that $M_T$ obeys 3D-xy scaling laws neat $T_c$. We showed that $M_T$ exhibits similar behaviour as $C$ when crossing the vortex-melting line: a peak in the case of first-order transitions, and a step in the case of second-order transitions. Moreover, the parameters describing the anomalies, such as the latent heat and the step height, fulfil requirements of thermodynamic consistency, establishing the isothermal magnetocaloric effect as a useful tool for studying reversible processes in a superconductor.

At lower temperature the irreversible contribution to $M_T$ which results from the heat released by the friction of vortices moving across pinning centers becomes important. The hysteresis $\Delta M_T$ is similar to that of the magnetization, and is also a measure of the critical current density $J_c$. In this regime, characteristic features of the magnetization such as the fishtail effect are reflected in magnetocaloric runs. The phase diagram (figure 12) shows that the irreversibility lines obtained from $M(B_a)$ and $M_T(B_a)$ can be easily connected together, yielding a common irreversibility line which announces the fishtail effect. The latter is clearly observed at lower temperature in the form of a valley in the $J_c$ distribution.

Loosely speaking, one might conclude that in the reversible region $M_T$ behaves more like $C$ (figure 9), whereas in the irreversible region $M_T$ behaves more like $M$ (figure 10). Thus, using a single experimental setup for the isothermal magnetocaloric effect it is possible to investigate both reversible and irreversible phenomena in a superconductor. The results of magnetocaloric investigations for a high quality detwinned Y-123 single crystal will be published in a forthcoming paper.

## 6. Acknowledgements

This work was supported by the Swiss National Science Foundation through the national Centre of Competence in Research "Material with Novel Electronic Properties - MaNEP" and the Polish State Committee for Scientific Research under contract No. 2 POB 036 24.